\documentclass[11pt,twoside]{article}
\usepackage{asp2010}

\resetcounters

\begin{document}

\title{The Palomar Transient Factory: High Quality Realtime Data Processing in a Cost-Constrained Environment}
\author{Jason~Surace$^1$, Russ Laher$^1$, Frank~Masci$^1$, Carl~Grillmair$^1$, George~Helou$^1$
\affil{$^1$Infrared Processing and Analysis Center, California Institute of Technology, Pasadena, CA, 91125}
}

\begin{abstract}
The Palomar Transient Factory (PTF) is a synoptic sky survey in operation since 2009. PTF utilizes a 7.1 square degree camera on the Palomar 48-inch Schmidt telescope to survey the sky primarily at a single wavelength (R-band) at a rate of 1000-3000 square degrees a night. The data are used to detect and study transient and moving objects such as gamma ray bursts, supernovae and asteroids, as well as variable phenomena such as quasars and Galactic stars. The data processing system at IPAC handles realtime processing and detection of transients, solar system object processing, high photometric precision processing and light curve generation, and long-term archiving and curation. This was developed under an extremely limited budget profile in an unusually agile development environment. Here we discuss the mechanics of this system and our overall development approach.

Although a significant scientific installation in of itself, PTF also serves as the prototype for our next generation project, the Zwicky Transient Facility (ZTF). Beginning operations in 2017, ZTF will feature a 50 square degree camera which will enable scanning of the entire northern visible sky every night. ZTF in turn will serve as a stepping stone to the Large Synoptic Survey Telescope (LSST), a major NSF facility scheduled to begin operations in the early 2020s.

\end{abstract}

\section{The Palomar Transient Factory}

The Palomar Transient Factory evolved from ideas germinating in the survey work of the 1990s.  This included numerous large areal ground-based optical surveys (e.g. the SDSS, CFHT-LS, etc), several of which contained long-timescale synoptic components. Other experiments centered specifically on supernova detection, such as the High-z SNe Search and the Supernova Cosmology Project. At the same time, a push developed for asteroid threat detection including a mandate from the US government. Several projects were developed specifically focussing on time-variable, transient science including LINEAR, Palomar Quest, the CRTS, and PAN-STARRS.  
Interest developed at Caltech for a survey that was structured specifically around various cadences aimed at SNe on scales larger than had previously been attempted. Advancements in computer technology allowed for faster processing, including turnaround within the same night.
PTF was jumpstarted with initial seed money from several sources. The original PTF partners were Caltech, LBNL, IPAC, U.C. Berkeley, LCO-GT, Oxford, Columbia, and the Weizmann Institute of Israel.  Although many of these are no longer members of the current collaboration, they should be recognized here. In particular, critical contributions to the early success of PTF were made by U.C. Berkeley (the machine learning layer) and LBNL/NERSC (the initial image subtraction and transient detection pipeline).

In 2013 the original PTF partnership expired and a new collaboration was formed. Known as the ``intermediate'' Palomar Transient Factory (or iPTF), this is intended to bridge the operational gap between PTF and the start of our next generation project. The new partners are Caltech, IPAC, the IPMU Kavli Insitute of Tokyo, Los Alamos National Laboratory, the Oskar Klein Centre of Stockholm, the University of Wisconsin-Madison, the University System of Taiwan, and the Weizmann Institute. iPTF differs from PTF primarily in the mode of science operations. Whereas PTF concentrated on several interlocking science programs with different cadences that were performed every night for several years, iPTF concentrates on specific focussed science campaigns, rotated on a quarterly basis. These campaigns allow observing cadences and targets that might not lend themselves well to timesharing with other programs.

The Infrared Processing and Analysis Center (IPAC) is a multi-mission science center located on the campus of the California Institute of Technology. While originally the NASA IRAS science center, it has evolved to handle a variety of space-based (e.g. ISO, Spitzer, Herschel, Planck, WISE) as well as ground-based projects (e.g. 2MASS, LSST, PTI, KOA, LCO-GT). IPAC also hosts several major archives, including the Infrared Science Archive (IRSA), the NASA Extragalactic Database (NED), and the NASA Exoplanet Database. IPAC has long history of experience with development of data processing pipelines and archive interfaces. Combined with its LSST involvement (science user interface), IPAC was a clear fit with PTF in terms of science and software development interest.

\vskip 0.1in
\noindent More information about PTF, iPTF, and ZTF can be found at \url{http://ptf.caltech.edu}.

\subsection{Data Acquisition System}

A substantial cost savings was achieved by retasking an existing camera (the CFHT 12k Mosaic Camera, \citet{2000SPIE.3965...58S}) rather than building an entirely new one.  The camera was substantially reworked \citep{2009PASP..121.1395L}, the most important alteration being replacement of the liquid nitrogen dewar with a mechanical cryo-cooler. One of the 12 CCDs was inoperable, but the cost risk for repair was too great.
While cosmetically undesirable, in terms of the primary project science goal (detecting transients) this had less impact than might be expected, and the survey mapping strategy results in coverage over these regions.
The camera also received a new field corrector, as well as a dry gas feed to prevent condensation from forming on the dewar window.
The camera was fitted with a sliding filter exchanger capable of holding two filters at any one time (nominally these are g and R-band, with an H$\alpha$ filter for use during the full moon). 
The vast majority of the survey is executed at R-band. This use of a single filter for the majority of the survey was deliberate. Since the 48-inch PTF telescope is intended primarily as a discovery engine, the use of a single filter maximizes the swept discovery volume (and hence cadence).
A sliding shutter was also installed in front of the camera.

\begin{figure}[!ht]
\centering
\includegraphics[width=4.5in]{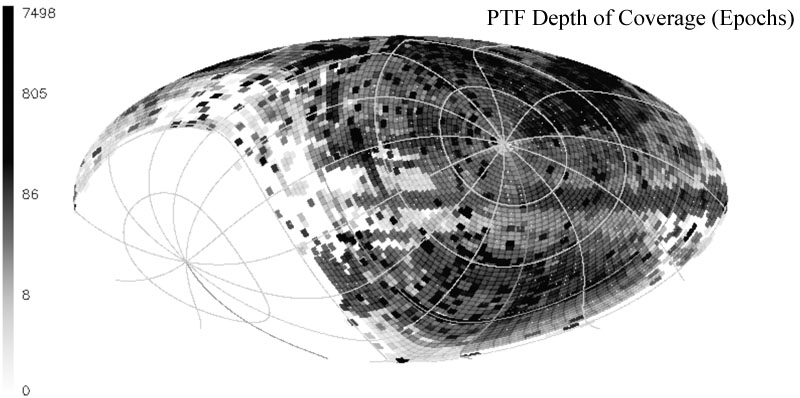}
\caption{Number of independent observations at R-band per location on the sky over the current 6-year lifespan of the PTF/iPTF project. This is an aitoff projection of the sky in galactic coordinates (i.e. the galactic plane extends horizontally across the image). Coordinate lines are celestial Right Ascension and Declination.}
\end{figure}

The telescope is the 48-inch (1.2m) Oschin-Schmidt telescope at Palomar. By using a pre-existing Caltech-owned facility, the cost savings was considerable over other projects that have constructed new telescopes. This is the same telescope used for the first ``big data'' project in astronomy, the original Palomar Observatory Sky Survey (POSS), as well as the later DPOSS and Quest projects. The  PTF camera is mounted inside the telescope tube in the location of the original photographic plate holder.

Telescope operations are fully robotic, saving the time expense and time inefficiencies associated with human observers. A human telescope operator at the nearby Hale 200-inch makes safety decisions about opening and closing the 48-inch dome. The robotic telescope controller autonomously follows a basic set of observation priorities, dynamically adapting observations to maintain the desired cadence. PTF does not follow a simple raster pattern on the sky. Rather, the observation pattern is science driven, with certain types of targets having different cadences. 

At large fraction of PTF science involves rapid transient response. New transient sources (SNe, asteroids) are discovered and scheduled for follow-up, sometimes {\it in the same night}, at the Palomar 60-inch, the 200-inch, LCO-GT, or other facilities. This integrated usage of the Palomar telescopes (discovery with the wide-field 48-inch with coordinated followup of bright sources with the 60-inch and faint sources with the 200-inch) actually mirrors the original Palomar design \citep{1972rsta.conf....5M}, only replacing the previous instruments with modern digital devices.

\subsection{Data Processing Layer (Operations) Physical Infrastructure}

Raw data is transmitted from Palomar via a microwave link, relaying to IPAC through the San Diego Supercomputing Center. Although currently adequate, the maximum bandwidth of this link will be one of the rate-limiting factors for our next generation project (described below). This data transmission issue is common to many synoptic surveys, and remains an outstanding infrastructural issue. Data processing cannot occur on the mountaintop. All of the images in a synoptic survey are of value, and as a result the data ``reduction'' process actually increases the overall data volume by factors of several. In addition, the physical plant (space, power, cooling, etc) available at Palomar are inadequate to host the required computing resources on-site.

There are twenty-four compute drones, half of which are dual-cpu 8-core Sun systems, while the other half are dual-cpu 12-core Dells. All run Red Hat Enterprise linux. The fileserver and database servers are Sun systems running Solaris. A10G ethernet network is used. All of the drones were purchased new, although in some cases second-hand units were used for beta testing.

Disk storage is based on Nexsan ATABeast network attached storage. In addition, the individual compute drones are equipped with internal scratch disks. In most cases these were purchased new, although two of the units consisted of used chassis populated with new disks. The filesystem is ZFS. In-line filesystem data compression is used which generally recovers about 30\% of the physical storage.
An unusual feature of this system is that the long-term storage is dual-homed and connected to both the operations system and the archive system. This was a financial consideration: the spinning disk is too expensive to replicate both an operations and archive copy. This adds an additional operational complexity cost, since it requires careful control of file ownership between the operations and archive layers.

The PTF system lies behind the IPAC facility  two-factor authentication firewall. The only exposed components are the transfer computer used to receive data, and the archive interface computers. Because PTF shares facilities infrastructure with all of IPAC (including a state of the art data room with support personnel), there is an additional savings through economy of scale.

\subsection{Data Processing Layer (Operations) Software}

The data processing system draws on design heritage from numerous IPAC space missions.
The pipeline building blocks are individual modules that carry out specific data reduction operations (for example, deriving and removing the CCD bias pattern). These modules are written in a variety of languages, the most common being C, while others are in Python, IDL, and Matlab. The choice of language was generally dictated by the preference and proficiency of the individual coder. The interfaces to these modules are regularized through use of PERL wrappers. 
The pipeline makes heavy use of community software. These include SWARP, SCAMP, Sextractor, Daophot, and the astrometry.net astrometric calibration suite. Early versions of the image subtraction pipeline also used HOTPANTS. 

The pipeline executor is also written in PERL \citep{2014PASP..126..674L}. The system divides the major data reduction tasks into individual pipelines composed of multiple modules; a virtual pipeline operator handles job distribution and automated daily operations while a human pipeline operator chooses which additional tasks to queue.
A postgres database is central to pipeline operations and is used to track pipeline status, data flow, and  data quality metrics. Catalog (extracted source photometry) is {\it  not} loaded into a relational database within the operations system as early experiments showed this to be too costly in terms of performance, and flat file access is used instead.

The processing of PTF data is embarrassingly parallel.
The parallel quantization is to process all data based on spatial location. Specifically, all PTF observations occur on a fixed tiling scheme on the sky referred to as a ``ptffield id''. The combination of fieldid, CCD, and filter uniquely spatially describe all data. Processes based on ensembles of data are processed entirely within a drone given this spatial assortment.  Individual CCDs are sent to individual drones. Data is processed locally using the drone internal scratch disks. Periodically, data that needs to be shared is written back to ``sandbox'' disks, which are a large shared disk pool connected to all the drones. After all the data products are produced, they are then copied into long-term archive storage. A principle in processing systems is ``keep the data close to the cpus''. This is commonly enabled through application layers that manage data. In our case this is accomplished directly by the pipeline design and network. 

Because the data acquisition system is an autonomous robot, we do not have foreknowledge of the events in a given night and we only learn about the observations upon receipt of the data. This is in contrast to most space missions, which use a shared uplink/downlink database, and makes the staging of pipelines more complex. Future versions of this pipeline will address this issue.

\begin{figure}[!ht]
\centering
\includegraphics[width=4.0in]{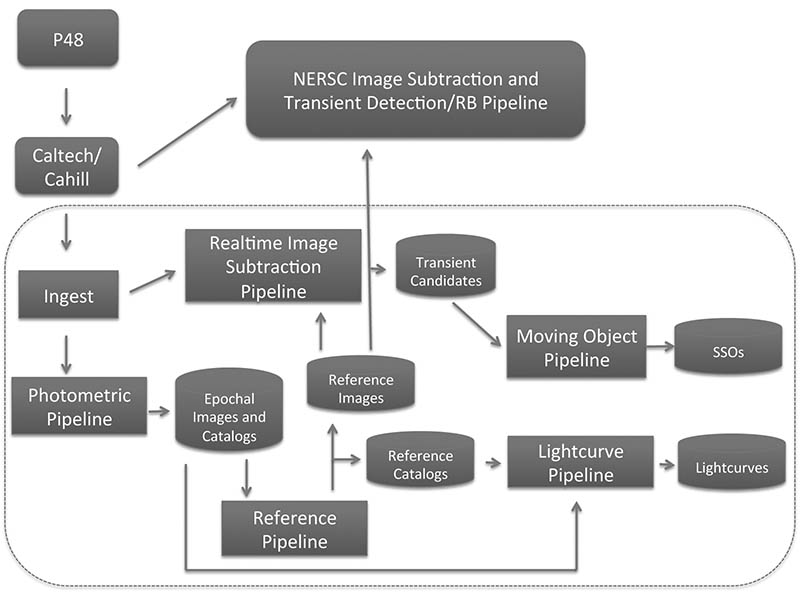}
\caption{Schematic flowchart of the iPTF data flow. The encircled area is processing that occurs within IPAC. The data flow has many interconnected branches with reliances between the segments on data products produced by other segments.}
\end{figure}

The primary pipeline segments are:

\begin{itemize}
\item Data Ingest --- The raw multi-extension FITS files are pushed into a transfer/receiving computer on the border of the IPAC firewall. The files are registered in the database and placed into the raw data archive,  and an initial astrometric calibration is made. At this stage various checks are made for inconsistent metadata in the incoming data products. The files are then split apart into individual CCDs and directed to the various drones for processing.

\item Realtime Pipeline --- As soon as data arrive, a realtime pipeline is triggered. This is primarily an image-subtraction pipeline. The data go through an initial calibration (bias subtraction, flat-fielding) using pre-existing calibration files from a previous night. A refined photometric calibration is derived. Software developed in-house provides photometric, astrometric,  and beam-shape matching between the incoming frame and the deeper reference images, producing a frame where all of the static objects have been subtracted out.  In-house code is used for beam-matching implementing the Bramich algorithm \citep{2008MNRAS.386L..77B}.
A source detector is run, creating transient candidates. These candidates are then further machine-vetted by a machine-learning algorithm called RealBogus, developed originally at U.C. Berkeley and now under further development by JPL and LANL. The resulting candidate lists are then sent to the extragalactic marshall as likely real transients. In addition, a ``streak detector'', also developed in-house (originally for detecting aircraft and satellite trails), is used to detect extremely fast-moving solar system objects. These are also vetted by a machine-learning algorithm. All of these candidates are then directed to a moving object pipeline, which constructs potential tracklets across multiple observations, resulting in lists of solar system objects.
This pipeline currently completes processing roughly 20 minutes after the data are taken. However, because jobs are stacked, this is a constant phase lag and the processing keeps up with the data flow.

\item Photometric Pipeline --- When the last observations for a given night are sent, they include an ``end of night'' signal indicating that no further data is expected. At this point a pipeline triggers which is intended to produce the highest quality data products possible, mirroring the usual data reduction steps used by astronomers when hand-reducing data. Specifically, data from the entire night is used to construct the best possible calibration for the flatfield and bias response, and to model the photometric calibration to compensate for changing atmospheric conditions. The astrometric calibration is tied to a combined UCAC4-SDSS catalog, and is presented in both the PV and TAN-SIP representations. The photometric calibration is tied to the SDSS, using observations of all SDSS-overlapping fields throughout the night, and in-house software to propagate the calibration to non-SDSS fields. Along with the reduced images, matching bit-masks are created which encode information such as radiation hits, CCD bleeds, saturation, and filter ghosts. Source catalogs are generated both with aperture photometry (tuned for galaxies) and psf-fitting (ideal for stars). The resulting images and catalogs are generally flux-calibrated to 2-3\% accuracy, with astrometric accuracy of 0.15\arcsec.

\item Reference Image Pipeline --- This pipeline generates deep images of the sky (called ``reference images''), which in turn are the basis for the image subtraction stage in the realtime pipeline. The inputs are the images produced by the photometric pipeline. Various checks are made to ensure that only the highest quality data gets coadded, and at present between 5 and 50 frames are used for input. A new photometric and astrometric calibration is derived in which the input frames are specifically aligned to each other, which is more accurate than relying on alignment through a secondary catalog. In-house software is used to derive the coaddition weights and perform the actual stacking. Deep reference catalogs (used by the final pipeline layer) are extracted using both aperture photometry and psf-fitting. These form the basis of the catalog of the static sky. As discussed, these reference images are made on a tile, chip, and filter basis. This point is important - the resulting reference images are observed with the same instrument and filter, pointed in nearly the same location on the sky. This minimizes issues during image subtraction with matching across effective instruments.

\item Light Curve Pipeline --- This pipeline connects the individual epochal measurements with actual astrophysical objects, and then refines the photometry to produce high accuracy light curves. Each night, as data are collected, a source matching pipeline connects each detection of each object based on spatial proximity to a source located in the deeper reference catalogs.  This approach is more successful than a generalized direct cross-match between epochal catalogs. This is because variations in spatial resolution from the atmosphere create ambiguities in the cross-matching, which in a generalized catalog approach quickly develops an exponentially large number of potential linkages. Once the cross-matches are calculated, roughly once a week or month photometric corrections are generated. It is assumed that the magnitude zeropoints for each image are in error by small amounts, presumably due to uncertainties in the propagation of the SDSS derived magnitudes, non-photometric conditions, etc. For each tile, CCD, and filter combination an analysis is performed to find the objects within the field with the lowest variances. These are then assumed to be static, and used to anchor the calibration for each individual image. This results in a table of delta-corrections that can be applied to each point in a light curve, after cross-referencing the parent image. The end result are light curves with an accuracy of a few millimag, although these are dominated by poisson noise at magnitudes fainter than 17.

\end{itemize}

\subsection{Archive Layer (IRSA)}

The function of the archive layer is to provide a means of discovering data products produced by the operations layer, and to deliver those data products to whatever their end use might be.
The PTF and ZTF data archive layers were developed by the Infrared Science Archive (IRSA) group within IPAC. IRSA is a multimission science archive, providing data services for Spitzer, 2MASS,  WISE and others. IRSA has developed  common toolkits upon which these mission archive interfaces were built. Leveraging this capability allowed a relatively low cost archive to be developed quickly.

\begin{figure}[!ht]
\centering
\includegraphics[width=5in]{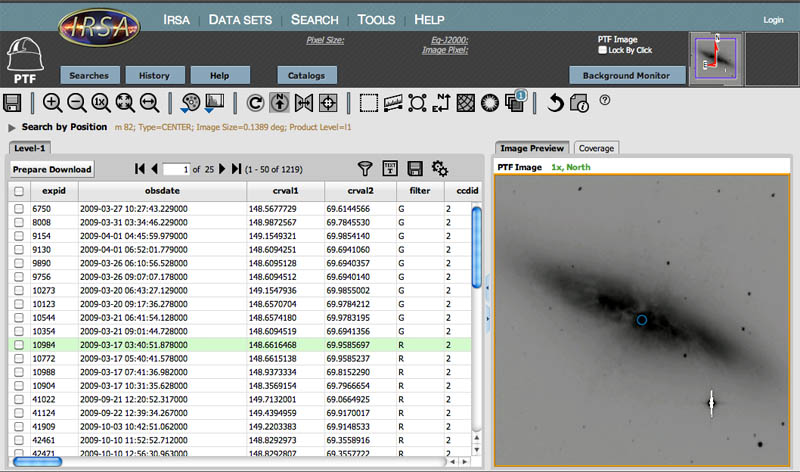}
\caption{PTF image search/discovery, exploration, and download tool at the Infrared Science Archive, based on a shared toolkit also used for Spitzer and WISE. Common usage pattern is for users to use the tool to rapidly find what sort of data holdings exist within the PTF archive.}
\end{figure}

After data is processed by the operations system at the end of each day, possession of the data is passed to IRSA. Image data is passed by file reference and ownership change. Catalog (database) data is passed via load files generated by the operations system, which are then ingested into an IRSA specific database. Several factors drive this separation between the archive and operations layers. The archive is intended as a reliable long-term storage system. This requires a guarantee that the files continue to exist on the shared disk system. Second, this allows the archive system to abstract the file storage media and paths, allowing transparent future hardware replacements and upgrades. Finally, the underlying storage schema required by the archive technology may currently be or will evolve differently from that of the operations system.

The archive currently offers several different services. The first of these is a GUI web-based data search tool and visualizer. Searches are enabled by name and position. In addition, there is a moving object based search that correlates an ephemeris against the position and time of the PTF observations. Images may be previewed through a live display pane. Both images and catalog files may be retrieved in bulk. The interface offers on-the-fly thumbnailing, since most users are interested in limited regions around their targets of interest.
A second interface is provided through a set of APIs that allow http-based calls (e.g. via wget or curl). This allows users to embed direct calls to the archive into their own software. Both this API and the web-based gui utilize the same underlying infrastructure, and a result offer similar services.

\subsection{Science Tools Layer (Marshals)}

A critical component of PTF for enabling rapid science results are the so-called ``science marshals''.
The marshals are organized around scientific topics, and serve many different functions which are unique to carrying out science analysis. In this sense, they are somewhat similar to the envisioned LSST ``event brokers''. One of the most important of these is the extragalactic marshall. It presents data from the realtime image subtraction and transient detection pipeline, including thumbnails of the target, image differences, and images from other surveys (e.g. SDSS) when available. Data fusion functions include automated cross-referencing to other catalogs, as well as historical retrieval of photometry from the region of interest. Most importantly, there are interactive functions allowing scientists to comment on the results, coordinate and upload follow-up observations, and additionally tag the data with new results.
A similar galactic marshall exists for analysis of light curve data and includes functions such as phase-wrapping of light curves on demand, while a solar system object marshall presents candidate fast-moving objects and shows prepared data for the Minor Planet Center. In most cases the marshalls incorporate the data archive as services, calling as needed data requests for thumbnails, etc.

\section{The Zwicky Transient Facility}

We are beginning construction of a new phase in the Palomar transient universe investigation: the Zwicky Transient Facility. At the heart of this program is a new camera with a field of view of nearly 50 square degrees. The enabling technology for this camera is the development of inexpensive ``wafer-scale'' CCDs, e.g. detectors the full size of silicon wafers. The ZTF camera will utilize 16 6kx6k CCDs manufactured by e2v. Through the use of parallel readouts and modern CCD control electronics, these devices can be fully read in roughly 10 seconds. Currently, we plan to utilize a 30-second exposure time, with focal plane readout occurring during the telescope slew and settle. This will allow ZTF to cover 3$\pi$ in 8 hours, or put more simply this will allow ZTF to image the entire visible northern sky every night. This will lead to a variety of new science capabilities. We highlight here one of the more interesting: gravity wave localization. Although gravity waves have yet to be detected, it has long been acknowledged that the existing detectors (e.g. LIGO) can only localize events to large areas (hundreds of square degrees) on the sky. When first proposed, this was a significant criticism; it would be impossible to connect gravity wave sources to any other observations or indeed to any specific astronomical sources on the sky. With ZTF's large areal coverage, depth, and fast turnaround time, it will be possible to survey the entire probable region of the gravity wave event in order to catch any rapidly fading electromagnetic counterpart, a feat iPTF has already demonstrated with gamma ray burst sources.

In addition to the new camera, ZTF will also require modest telescope modifications. The most notable of these is the use of a bi-parting shutter external to the telescope tube. Because the camera mechanism is located inside the telescope, it is extremely important that it have as small a cross-section as possible, and this precludes the use of a focal plane shutter. Similarly, a filter exchanger is planned that will use a folding truss mechanism to stow the (2 of 3) filters not currently in use against the inner surface of the telescope. The camera itself has a remarkably small cross-section, being only slightly larger than the CCDs themselves. The  camera itself is very thin (being cooled via an external cryo-cooler) at roughly 4 inches thick.  Guiding, as well as tip/tilt and focus measurements, are provided by four 2048x2048 CCDs, each of which is nearly a quarter degree by themselves.

ZTF began receiving funding from the U.S. National Science Foundation (NSF) starting fiscal year 2015 under the Mid-Scale Innovations Program (MSIP). The MSIP award, together with contributions from private partners, is sufficient to enable full funding of the project through completion of three years of operations in 2020. The instrument is expected to see first light at the start of 2017. All data will be publicly released. This will start with the PTF image and catalog data in fall of 2015, followed by a light curve release in the following year. The first year of ZTF operations will be devoted to reference image building. Eventually, all catalog and image data as well as transient alerts will be released on an annual basis. 

ZTF is the most immediate predecessor to the Large Synoptic Survey Telescope (LSST), a major NSF-funded facility scheduled for first light in 2019 and science operations beginning around 2022. ZTF will possess a similar cadence to LSST, but the latter will have superior image quality, multiple simultaneous filters, and will be much deeper as the telescope is more than 6 times larger in radius. Nonetheless, the two surveys are actually highly complementary - in particular, ZTF's prime discovery space will lie at magnitudes that are very bright for LSST. 

\section{System Development Under Severe Constraints}

PTF has been enormously successful despite an extremely limited monetary budget. Most IPAC projects are NASA space missions, with substantially larger budgets and which follow a classic ``waterfall'' development approach: substantial effort devoted to science definition and requirements writing, followed by technical design requirements, and finally coding and testing with fixed delivery cycles, typically on six month scales. Such an approach was inapplicable to PTF for several reasons. First, we lacked the funded manpower required for any effort beyond coding and testing. The core team consisted of roughly three people, with additional staff brought in as needed. Major contributions were also made by 3-5 graduate students and post-docs, as time permitted and the project goals aligned with their science goals. Second, we were on an accelerated timetable, where the software in its initial form had to be brought online in roughly one year, precluding a lengthy design phase. Finally, at the start of the project we were uncertain of the achievable data quality, or indeed what science goals could be accomplished, thus precluding writing very detailed science requirements. PTF therefore proceeded under a very ``agile'' development cycle.

A first critical element was that development  remained science focussed. Astronomy has become a highly competitive field, and the most common metric of success is paper production. Therefore, we allowed the evolving nature of the science possible with PTF to guide the direction of development. Astronomy shifts focus on 3-5 year timescales, so today's science drivers are unlikely to be tomorrow's.  A specific example has been the evolution of PTF away from being solely a transient discovery engine. As it became obvious that the system could support high accuracy light curves, the emphasis for software development has shifted away from SNe (of which there are now thousands spectroscopically confirmed) to variable phenomena like Galactic stars.

The development program also emphasized rapid results with a philosophy of establishing basic functionality and then addressing incremental improvements and adding capabilities as needed. 
Big data projects have data volumes that are, by definition, too large to be amenable to hand analysis.
The data system must be capable of rendering the data into a form sufficiently intelligible for further analysis and refinement; attempting to reach the end state during development is unlikely to succeed.
In order to speed development we made extremely abundant use of community software (see Section 1.3), and functionality was valued over elegance. While the system does not use the latest technology, it was brought partially on-line in under one year, and science papers were written almost immediately.

Finally we used an extremely agile development process.
Development was by a very small group. The coders and leads were all scientists. IPAC has always insisted on deep involvement by research astronomers in the development process; this ensures that the software produces results that will actually meet the needs of the science users. The use of students and post-docs also accelerates the development, since they have a vested interest in the software actually working.
In the long term this creates sustainability issues, but in the short term results in productive software at low cost and is a viable approach depending on the scope of the project. 

In conclusion, while our efforts were clearly defined in many ways by the limitations of our budget, PTF (and soon to be ZTF) are clear examples of a high science per dollar project made possible by properly aligning software development and science goals.

\end{document}